# MIScnn: A Framework for Medical Image Segmentation with Convolutional Neural Networks and Deep Learning

Dominik Müller[1] and Frank Kramer[1]

[1] IT-Infrastructure for Translational Medical Research, Faculty of Applied Computer Science, Faculty of Medicine, University of Augsburg, Germany

**Abstract.** The increased availability and usage of modern medical imaging induced a strong need for automatic medical image segmentation. Still, current image segmentation platforms do not provide the required functionalities for plain setup of medical image segmentation pipelines. Already implemented pipelines are commonly standalone software, optimized on a specific public data set. Therefore, this paper introduces the open-source Python library MIScnn. The aim of MIScnn is to provide an intuitive API allowing fast building of medical image segmentation pipelines including data I/O, preprocessing, data augmentation, patch-wise analysis, metrics, a library with state-of-the-art deep learning models and model utilization like training, prediction, as well as fully automatic evaluation (e.g. cross-validation). Similarly, high configurability and multiple open interfaces allow full pipeline customization. Running a cross-validation with MIScnn on the Kidney Tumor Segmentation Challenge 2019 data set (multi-class semantic segmentation with 300 CT scans) resulted into a powerful predictor based on the standard 3D U-Net model. With this experiment, we could show that the MIScnn framework enables researchers to rapidly set up a complete medical image segmentation pipeline by using just a few lines of code. The source code for MIScnn is available in the Git repository: *https://github.com/frankkramer-lab/MIScnn*.

**Keywords:** Medical image analysis; segmentation; computer aided diagnosis; biomedical image segmentation; u-net, deep learning; convolutional neural network; open-source; framework.

## 1 Introduction

Medical imaging became a standard in diagnosis and medical intervention for the visual representation of the functionality of organs and tissues. Through the increased availability and usage of modern medical imaging like Magnetic Resonance Imaging (MRI), the need for automated processing of scanned imaging data is quite strong [1]. Currently, the evaluation of medical images is a manual process performed by physicians. Larger numbers of slices require the inspection of even more image material by doctors, especially regarding the increased usage of high-resolution medical imaging. In order to shorten the time-consuming inspection and evaluation process, an automatic pre-segmentation of abnormal features in medical images would be required.

Image segmentation is a popular sub-field of image processing within computer science [2]. The aim of semantic segmentation is to identify common features in an input image by learning and then labeling each pixel in an image with a class (e.g. background, kidney or tumor). There is a wide range of algorithms to solve segmentation problems. However, state-of-the-art accuracy was accomplished by convolutional neural networks and deep learning models [2–6], which are used extensively today. Furthermore, the newest convolutional neural networks are able to exploit local and global features in images [7–9] and they can be trained to use 3D image information as well [10,11]. In recent years, medical image segmentation models with a convolutional neural network architecture have become quite powerful and achieved similar results performance-wise as radiologists [5,12]. Nevertheless, these models have been standalone applications with optimized architectures, preprocessing procedures, data augmentations and metrics specific for their data set and corresponding segmentation problem [9]. Also, the performance of such optimized pipelines varies drastically between different medical conditions. However, even for the same medical condition, evaluation and comparisons of these models are a persistent challenge due to the variety of the size, shape, localization and distinctness of different data sets. In order to objectively compare two segmentation model architectures from the sea of one-use standalone pipelines, each specific for a single public data set, it would be required to implement a complete custom pipeline with preprocessing, data augmentation and batch creation. Frameworks for general image segmentation pipeline building can not be fully utilized. The reason for this are their missing medical image I/O interfaces, their preprocessing methods, as well as their lack of handling highly unbalanced class distributions, which is standard in medical imaging. Recently developed medical image segmentation platforms, like NiftyNet [13], are powerful tools and an excellent first step for standardized medical image segmentation pipelines. However, they are designed more like configurable software instead of frameworks. They lack modular pipeline blocks to offer researchers the opportunity for easy customization and to help developing their own software for their specific segmentation problems.

In this work, we push towards constructing an intuitive and easy-to-use framework for fast setup of state-of-the-art convolutional neural network and deep learning models for medical image segmentation. The aim of our framework MIScnn (**M**edical **I**mage **S**egmentation with **C**onvolutional **N**eural **N**etworks) is to provide a complete pipeline for preprocessing, data augmentation, patch slicing and batch creation steps in order to start straightforward with training and predicting on diverse medical imaging data. Instead of being fixated on one model architecture, MIScnn allows not only fast switching between multiple modern convolutional neural network models, but it also provides the possibility to easily add custom model architectures. Additionally, it facilitates a simple deployment and fast usage of new deep learning models for medical image segmentation. Still, MIScnn is highly configurable to adjust hyperparameters, general training parameters, preprocessing procedures, as well as include or exclude data augmentations and evaluation techniques.

## 2 Methods

The open-source Python library MIScnn is a framework to setup medical image segmentation pipelines with convolutional neural networks and deep learning models. MIScnn is providing several core features:

- 2D/3D medical image segmentation for binary and multi-class problems
- Data I/O, preprocessing and data augmentation for biomedical images
- Patch-wise and full image analysis
- State-of-the-art deep learning model and metric library
- Intuitive and fast model utilization (training, prediction)
- Multiple automatic evaluation techniques (e.g. cross-validation)
- Custom model, data I/O, pre-/postprocessing and metric support

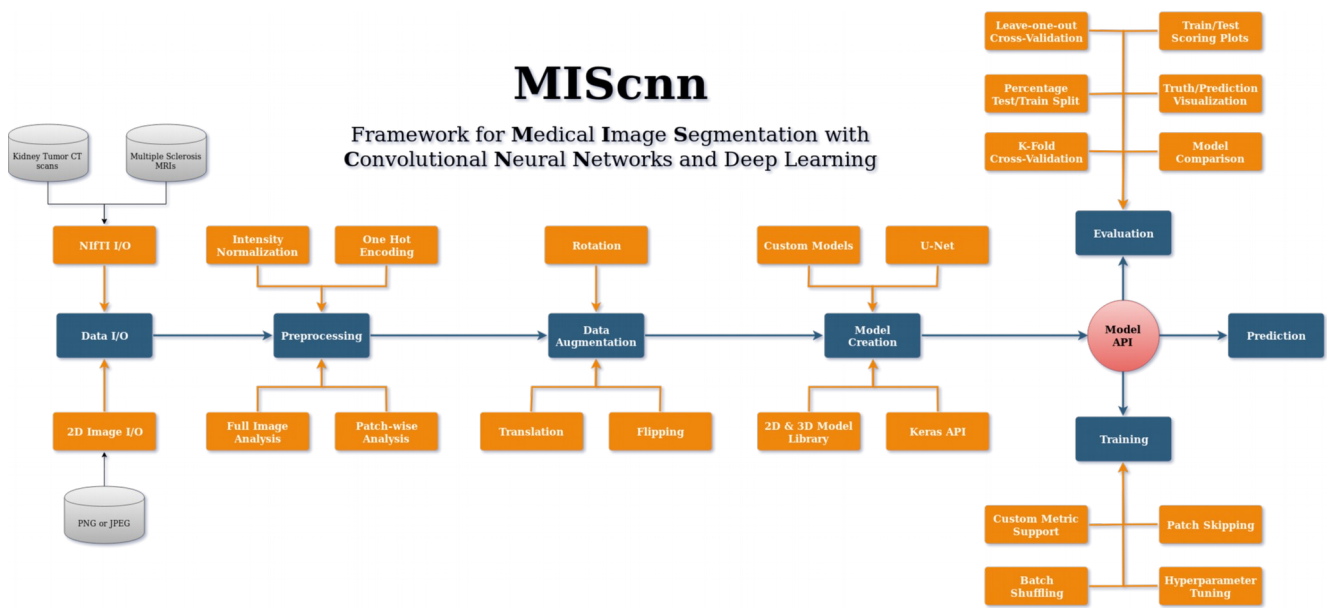

**Figure 1:** Flowchart diagram of the MIScnn pipeline starting with the data I/O and ending with a deep learning model.

### 2.1 Data Input

*NIfTI Data I/O Interface*

MIScnn provides a data I/O interface for the Neuroimaging Informatics Technology Initiative (NifTI) [14] file format for loading Magnetic Resonance Imaging and Computed Tomography data into the framework. This format was initially created to speed up the development and enhance the utility of informatics tools related to neuroimaging. Still, it is now commonly used for sharing public and anonymous MRI and CT data sets, not only for brain imaging, but also for all kinds of human 3D imaging. A NIfTI file contains the 3D image matrix and diverse metadata, like the thickness of the MRI slices.

*Custom Data I/O Interface*

Next to the implemented NIfTI I/O interface, MIScnn allows the usage of custom data I/O interfaces for other imaging data formats. This open interface enables MIScnn to handle specific biomedical imaging features (e.g. MRI slice thickness), and therefore it avoids losing these feature information by a format conversion requirement. A custom I/O interface must be committed to the preprocessing function and it has to return the medical image as a 2D or 3D matrix for integration in the workflow. It is advised to add format specific preprocessing procedures (e.g. MRI slice thickness normalization) in the format specific I/O interface, before returning the image matrix into the pipeline.

## 2.2 Preprocessing

*Pixel Intensity Normalization*

Inconsistent signal intensity ranges of images can drastically influence the performance of segmentation methods [15,16]. The signal ranges of biomedical imaging data are highly varying between data sets due to different image formats, diverse hardware/instruments (e.g. different scanners), technical discrepancies, and simply biological variation [5]. Additionally, the machine learning algorithms behind image segmentation usually perform better on features which follow a normal distribution. In order to achieve dynamic signal intensity range consistency, it is advisable to scale and standardize imaging data. The signal intensity scaling projects the original value range to a predefined range usually between [0,1] or [-1,1], whereas standardization centers the values close to a normal distribution by computing a Z-Score normalization. MIScnn can be configured to include or exclude pixel intensity scaling or standardization on the medical imaging data in the pipeline.

*Clipping*

Similar to pixel intensity normalization, it is also common to clip pixel intensities to a certain range. Intensity values outside of this range will be clipped to the minimum or maximum range value. Especially in computer tomography images, pixel intensity values are expected to be identical for the same organs or tissue types even in different scanners [17]. This can be exploited through organ-specific pixel intensity clipping.

*Resampling*

The resampling technique is used to modify the width and/or height of images. This results into a new image with a modified number of pixels. Magnetic resonance or computer tomography scans can have different slice thickness. However, training neural network models requires the images to have the same slice thickness or voxel spacing. This can be accomplished through resampling. Additionally, downsampling images reduces the required GPU memory for training and prediction.

*One Hot Encoding*

MIScnn is able to handle binary (background/cancer) as well multi-class (background/kidney/liver/lungs) segmentation problems. The representation of a binary segmentation is being made quite simple by using a variable with two states, zero and one. But for the processing of multiple categorical segmentation labels in machine learning algorithms, like deep learning models, it is required to convert the classes into a more mathematical representation. This can be achieved with the One Hot encoding method by creating a single binary variable for each segmentation class. MIScnn automatically One Hot encodes segmentation labels with more than two classes.

## 2.2 Patch-wise and Full Image Analysis

Depending on the resolution of medical images, the available GPU hardware plays a large role in 3D segmentation analysis. Currently, it is not possible to fully fit high-resolution MRIs with an example size of 400x512x512 into state-of-the-art convolutional neural network models due to the enormous GPU memory requirements. Therefore, the 3D medical imaging data can be either sliced into smaller cuboid patches or analyzed slice-by-slice, similar to a set of 2D images [5,6,18]. In order to fully use the information of all three axis, MIScnn slices 3D medical images into patches with a configurable size (e.g. 128x128x128) by default. Depending on the model architecture, these patches can fit into GPUs with RAM sizes of 4GB to 24GB, which are commonly used in research. Nevertheless, the slice-by-slice 2D analysis, as well as the 3D patch analysis is supported and can be used in MIScnn. It is also possible to configure the usage of full 3D images in case of analyzing uncommonly small medical images or having a large GPU cluster. By default, 2D medical images are fitted completely into the convolutional neural network and deep learning models. Still, a 2D patch-wise approach for large resolution images can be also applied.

## 2.3 Data Augmentation for Training

In the machine learning field, data augmentation covers the artificially increase of training data. Especially in medical imaging, commonly only a small number of samples or images of a studied medical condition is available for training [5,19–22]. Thus, an image can be modified with multiple techniques, like shifting, to expand the number of plausible examples for training. The aim is to create reasonable variations of the desired pattern in order to avoid overfitting in small data sets [21].

For state-of-the-art data augmentation, MIScnn integrated the batchgenerators package from the Division of Medical Image Computing at the German Cancer Research Center (DKFZ) [23]. It offers various data augmentation techniques and was used by the winners of the latest medical image processing challenges [9,17,24]. It supports spatial translations, rotations, scaling, elastic deformations, brightness, contrast, gamma and noise augmentations like Gaussian noise.

## 2.4 Sampling and Batch generation

*Skipping Blank Patches*

The known problem in medical images of the large unbalance between the relevant segments and the background results into an extensive amount of parts purely labeled as background and without any learning information [5,19]. Especially after data augmentation, there is no benefit to multiply these blank parts or patches [25]. Therefore, in the patch-wise model training, all patches, which are completely labeled as background, can be excluded in order to avoid wasting time on unnecessary fitting.

*Batch Management*

After the data preprocessing and the optional data augmentation for training, sets of full images or patches are bundled into batches. One batch contains a number of prepared images which are processed in a single step by the model and GPU. Sequential for each batch or processing step, the neural network updates its internal weights accordingly with the predefined learning rate. The possible number of images inside a single batch highly depends on the available GPU memory and has to be configured properly in MIScnn. Every batch is saved to disk in order to allow fast repeated access during the training process. This approach drastically reduces the computing time due to the avoidance of unnecessary repeated preprocessing of the batches. Nevertheless, this approach is not ideal for extremely large data sets or for researchers without the required disk space. In order to bypass this problem, MIScnn also supports "on-the-fly" generation of the next batch in memory during runtime.

*Batch Shuffling*

During model training, the order of batches, which are going to be fitted and processed, is shuffled at the end of each epoch. This method reduces the variance of the neural network during fitting over an epoch and lowers the risk of overfitting. Still, it must be noted, that only the processing sequence of the batches is shuffled and the data itself is not sorted into a new batch order.

*Multi-CPU and -GPU Support*

MIScnn also supports the usage of multiple GPUs and parallel CPU batch loading next to the GPU computing. Particularly, the storage of already prepared batches on disk enables a fast and parallelizable processing with CPU as well as GPU clusters by eliminating the risk of batch preprocessing bottlenecks.

## 2.5 Deep Learning Model Creation

*Model Architecture*

The selection of a deep learning or convolutional neural network model is the most important step in a medical image segmentation pipeline. There is a variety of model architectures and each has different strengths and weaknesses [7,8,10,11,23–29]. MIScnn features an open model interface to load and switch between provided state-of-the-art convolutional neural network models like the popular U-Net model [7]. Models are represented with the open-source neural network library Keras [33] which provides an user-friendly API for commonly used neural-network building blocks on top of TensorFlow [34]. The already implemented models are highly configurable by definable number of neurons, custom input sizes, optional dropout and batch normalization layers or enhanced architecture versions like the Optimized High Resolution Dense-U-Net model [10]. Additionally, MIScnn offers architectures for 3D, as well as 2D medical image segmentation. Besides the flexibility in switching between already implemented models, the open model interface enables the ability for custom deep learning model implementations and simple integrating these custom models into the MIScnn pipeline.

*Metrics*

MIScnn offers a large quantity of various metrics which can be used as loss function for training or for evaluation in figures and manual performance analysis. The Dice coefficient, also known as the Dice similarity index, is one of the most popular metrics for medical image segmentation. It scores the similarity between the predicted segmentation and the ground truth. However, it also penalizes false positives comparable to the precision metric. Depending on the segmentation classes (binary or multi-class), there is a simple and class-wise Dice coefficient implementation. Whereas the simple implementation only accumulates the overall number of correct and false predictions, the class-wise implementation accounts the prediction performance for each segmentation class which is strongly recommended for commonly class-unbalanced medical images. Another popular supported metric is the Jaccard Index. Even though it is similar to the Dice coefficient, it does not only emphasize on precise segmentation. However, it also penalizes under- and over-segmentation. Still, MIScnn uses the Tversky loss [35] for training. Comparable to the Dice coefficient, the Tversky loss function addresses data imbalance. Even so, it achieves a much better trade-off between precision and recall. Thus, the Tversky loss function ensures good performance on binary, as well as multi-class segmentation. Additionally, all standard metrics which are included in Keras, like accuracy or cross-entropy, can be used in MIScnn. Next to the already implemented metrics or loss functions, MIScnn offers the integration of custom metrics for training and evaluation. A custom metric can be implemented as defined in Keras, and simply be passed to the deep learning model.

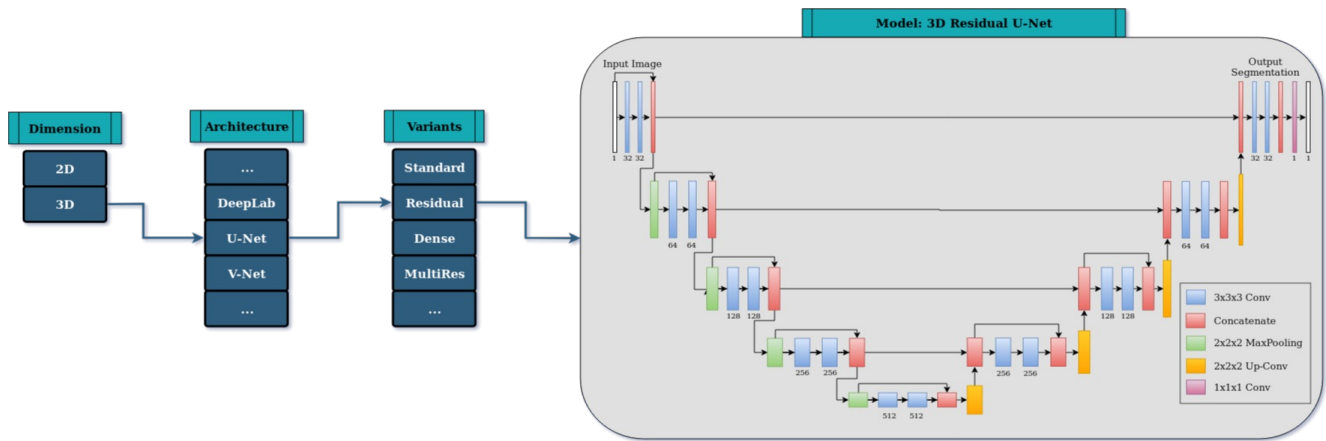

**Figure 2:** Flowchart visualization of the deep learning model creation and architecture selection process.

### 2.6 Model Utilization

With the initialized deep learning model and the fully preprocessed data, the model can now be used for training on the data to fit model weights or for prediction by using an already fitted model. Alternatively, the model can perform an evaluation, as well, by running a cross-validation for example, with multiple training and prediction calls. The model API allows saving and loading models in order to subsequently reuse already fitted models for prediction or for sharing pre-trained models.

*Training*

In the process of training a convolutional neural network or deep learning model, diverse settings have to be configured. At this point in the pipeline, the data augmentation options of the data set, which have a large influence on the training in medical image segmentation, must be already defined. Sequentially, the batch management configuration covered the settings for the batch size, and also the batch shuffling at the end of each epoch. Therefore, only the learning rate and the number of epochs are required to be adjusted before running the training process. The learning rate of a neural network model is defined as the extend in which the old weights of the neural network model are updated in each iteration or epoch. In contrast, the number of epochs defines how many times the complete data set will be fitted into the model. Sequentially, the resulting fitted model can be saved to disk.

During the training, the underlying Keras framework gives insights into the current model performance with the predefined metrics, as well as the remaining fitting time. Additionally, MIScnn offers the usage of a fitting-evaluation callback functionality in which the fitting scores and metrics are stored into a tab separated file or directly plotted as a figure.

*Prediction*

For the segmentation prediction, an already fitted neural network model can be directly used after training or it can be loaded from file. The model predicts for every pixel a Sigmoid value for each class. The Sigmoid value represents a probability estimation of this pixel for the associated label. Sequentially, the argmax of the One Hot encoded class are identified for multi-class segmentation problems and then converted back to a single result variable containing the class with the highest Sigmoid value.

When using the overlapping patch-wise analysis approach during the training, MIScnn supports two methods for patches in the prediction. Either the prediction process plainly creates distinct patches and treats the overlapping patches during the training as purely data augmentation, or overlapping patches are created for prediction. Due to the lack of prediction power at patch edges, computing a second prediction for edge pixels in patches, by using an overlap, is a commonly used approach. In the following merge of patches back to the original medical image shape, a merging strategy for the pixels is required, in the overlapping part of two patches and with multiple predictions. By default, MIScnn calculates the mean between the predicted Sigmoid values for each class in every overlapping pixel.

The resulting image matrix with the segmentation prediction, which has the identical shape as the original medical image, is saved into a file structure according to the provided data I/O interface. By default, using the NIfTI data I/O interface, the predicted segmentation matrix is saved in NIfTI format without any additional metadata.

*Evaluation*

MIScnn supports multiple automatic evaluation techniques to investigate medical image segmentation performance: k-fold cross-validation, leave-one-out cross-validation, percentage-split validation (data set split into test and train set with a given percentage) and detailed validation in which it can be specified which images should be used for training and testing. Except for the detailed validation, all other evaluation techniques use random sampling to create training and testing data sets. During the evaluation, the predefined metrics and loss function for the model are automatically plotted in figures and saved in tab separated files for possible further analysis. Next to the performance metrics, the pixel value range and segmentation class frequency of medical images can be analyzed in the MIScnn evaluation. Also, the resulting prediction can be compared directly next to the ground truth by creation image visualizations with segmentation overlays. For 3D images, like MRIs, the slices with the segmentation overlays are automatically visualized in the Graphics Interchange Format (GIF).

## 3 Experiments and Results

Here, we analyze and evaluate data from the Kidney Tumor Segmentation Challenge 2019 using MIScnn. All results were obtained using the scripts shown in the appendix.

### 3.1 Kidney Tumor Segmentation Challenge 2019 (KiTS19)

With more than 400 000 kidney cancer diagnoses worldwide in 2018, kidney cancer is under the top 10 most common cancer types in men and under the top 15 in woman [36]. The development of advanced tumor visualization techniques is highly important for efficient surgical planning. Through the variety in kidney and kidney tumor morphology, the automatic image segmentation is challenging but of great interest [24].

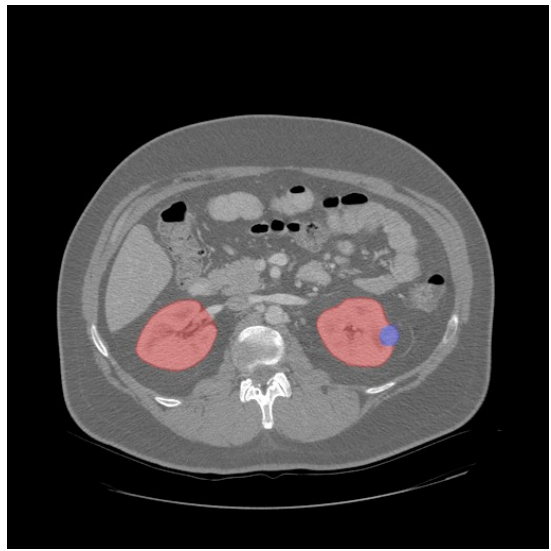

**Figure 3:** Computed Tomography image from the Kidney Tumor Segmentation Challenge 2019 data set showing the ground truth segmentation of both kidneys (red) and tumor (blue).

The goal of the KiTS19 challenge is the development of reliable and unbiased kidney and kidney tumor semantic segmentation methods [24]. Therefore, the challenge built a data set for arterial phase abdominal CT scan of 300 kidney cancer patients [24]. The original scans have an image resolution of 512x512 and on average 216 slices (highest slice number is 1059). For all CT scans, a ground truth semantic segmentation was created by experts. This semantic segmentation labeled each pixel with one of three classes: Background, kidney or tumor. 210 of these CT scans with the ground truth segmentation were published during the training phase of the challenge, whereas 90 CT scans without published ground truth were released afterwards in the submission phase. The submitted user predictions for these 90 CT scans will be objectively evaluated and the user models ranked according to their performance. The CT scans were provided in NIfTI format in original resolution and also in interpolated resolution with slice thickness normalization.

**3.2 Validation on the KiTS19 data set with MIScnn**

For the evaluation of the MIScnn framework usability and data augmentation quality, a subset of 120 CT scans with slice thickness normalization were retrieved from the KiTS19 data set. An automatic 3-fold cross-validation was run on this KiTS19 subset with MIScnn.

*MIScnn Configurations*

The MIScnn pipeline was configured to perform a multi-class, patch-wise analysis with 80x160x160 patches and a batch size of 2. The pixel value normalization by Z-Score, clipping to the range -79 and 304, as well as resampling to the voxel spacing 3.22x1.62x1.62.
For data augmentation, all implemented techniques were used. This includes creating patches through random cropping, scaling, rotations, elastic deformations, mirroring, brightness, contrast, gamma and Gaussian noise augmentations. For prediction, overlapping patches were created with an overlap size of 40x80x80 in x,y,z directions. The standard 3D U-Net with batch normalization layers were used as deep learning and convolutional neural network model. The training was performed using the Tversky loss for 1000 epochs with a starting learning rate of 1E-4 and batch shuffling after each epoch. The cross-validation was run on two Nvidia Quadro P6000 (24GB memory each), using 48GB memory and taking 58 hours.

*Results*

With the MIScnn pipeline, it was possible to successfully set up a complete, working medical image multi-class segmentation pipeline. The 3-fold cross-validation of 120 CT scans for kidney and tumor segmentation were evaluated through several metrics: Tversky loss, soft Dice coefficient, class-wise Dice coefficient, as well as the sum of categorical cross-entropy and soft Dice coefficient. These scores were computed during the fitting itself, as well as for the prediction with the fitted model. For each cross-validation fold, the training and predictions scores are visualized in figure 4 and sum up in table 1.

**Table 1:** Performance results of the 3-fold cross-validation for tumor and kidney segmentation with a standard 3D U-Net model on 120 CT scans from the KiTS19 challenge. Each metric is computed between the provided ground truth and our model prediction and then averaged between the three folds.

| Metric | Training | Validation |
|---|---|---|
| Tversky loss | 0.3672 | 0.4609 |
| Soft Dice similarity coefficient | 0.8776 | 0.8235 |
| Categorical cross-entropy | - 0.8584 | - 0.7899 |
| Dice similarity coefficient: Background | - | 0.9994 |
| Dice similarity coefficient: Kidney | - | 0.9319 |
| Dice similarity coefficient: Tumor | - | 0.6750 |

The fitted model achieved a very strong performance for kidney segmentation. The kidney Dice coefficient had a median around 0.9544. The tumor segmentation prediction showed a considerably good but weaker performance than the kidney with a median around 0.7912. Still, the predictive power is very impressive in the context of using only the standard U-Net architecture with mostly default hyperparameters. In the medical perspective, through the variety in kidney tumor morphology, which is one of the reasons for the KiTS19 challenge, the weaker tumor results are quite reasonable [24]. Also, the models were trained with only 38% of the original KiTS19 data set due to 80 images for training and 40 for testing were randomly selected. The remaining 90 CTs were excluded in order to reduce run time in the cross-validation. Nevertheless, it was possible to build a powerful pipeline for kidney tumor segmentation with MIScnn resulting into a model with high performance, which is directly comparable with modern, optimized, standalone pipeline [7,8,11,26].

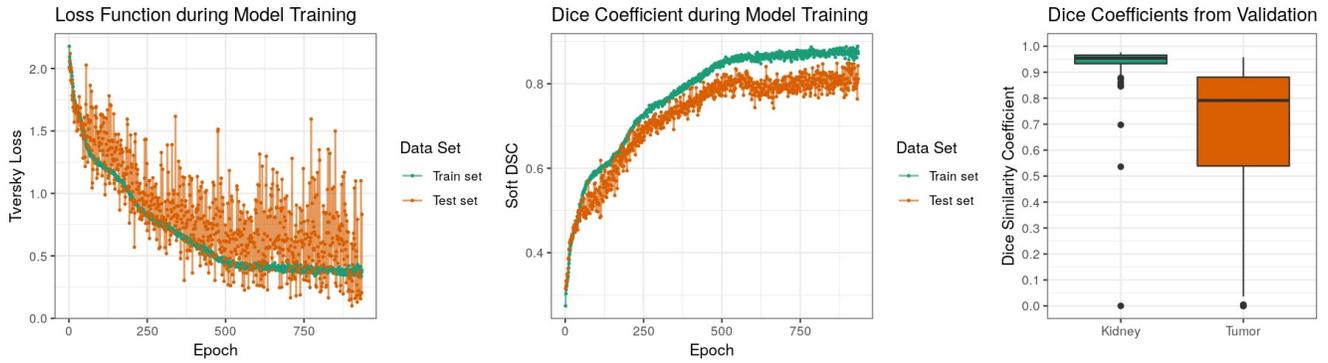

**Figure 4:** Performance evaluation of the standard 3D U-Net model for kidney and tumor prediction with a 3-fold cross-validation on the 120 CT data set from the KiTS19 challenge.

LEFT: Tversky loss against epochs illustrating loss development during training for the corresponding test and train data sets. Each point represents the average Tversky loss between the cross-validation folds.

CENTER: Class-wise Dice coefficient against epochs illustrating soft Dice similarity coefficient development during training for the corresponding test and train data sets. Each point represents the average soft Dice similarity coefficient between the cross-validation folds.

RIGHT: Dice similarity coefficient distribution for the kidney and tumor for all samples of the cross-validation.

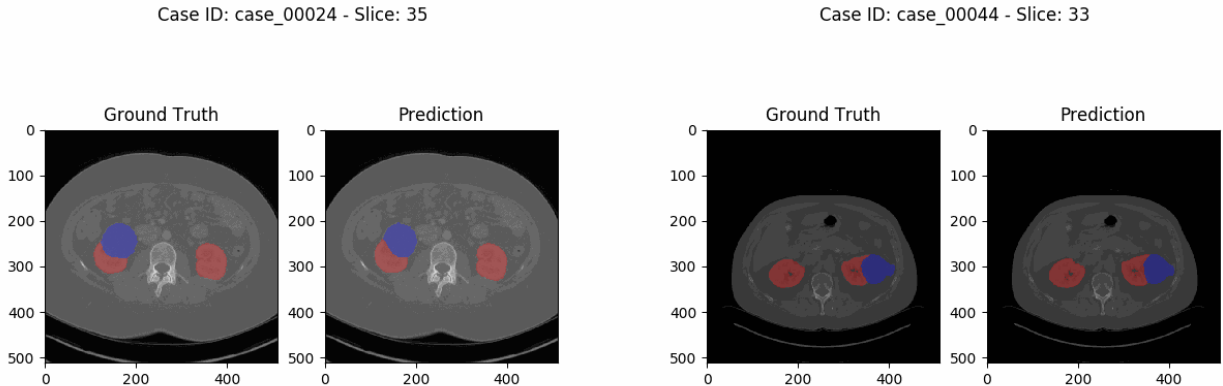

**Figure 5:** Computed Tomography scans of kidney tumors from the Kidney Tumor Segmentation Challenge 2019 data set showing the kidney (red) and tumor (blue) segmentation as overlays. The images show the segmentation differences between the ground truth provided by the KiTS19 challenge and the prediction from the standard 3D U-Net models of our 3-fold cross-validation.

Besides the computed metrics, MIScnn created segmentation visualizations for manual comparison between ground truth and prediction. As illustrated in figure 5, the predicted semantic segmentation of kidney and tumors is highly accurate.

## 4 Discussion

### 4.1 MIScnn framework

With excellent performing convolutional neural network and deep learning models like the U-Net, the urge to move automatic medical image segmentation from the research labs into practical application in clinics is uprising. Still, the landscape of standalone pipelines of top performing models, designed only for a single specific public data set, handicaps this progress. The goal of MIScnn is to provide a high-level API to setup a medical image segmentation pipeline with preprocessing, data augmentation, model architecture selection and model utilization. MIScnn offers a highly configurable and open-source pipeline with several interfaces for custom deep learning models, image formats or fitting metrics. The modular structure of MIScnn allows a medical image segmentation novice to setup a functional pipeline for a custom data set in just a few lines of code. Additionally, switchable models and an automatic evaluation functionality allow robust and unbiased comparisons between deep learning models. An universal framework for medical image segmentation, following the Python philosophy of simple and intuitive modules, is an important step in contributing to practical application development.

## 4.2 Use Case: Kidney Tumor Segmentation Challenge

In order to show the reliability of MIScnn, a pipeline was setup for kidney tumors segmentation on a CT image data set. The popular and state-of-the-art standard U-Net were used as deep learning model with up-to-date data augmentation. Only a subset of the original KiTS19 data set were used to reduce run time of the cross-validation. The resulting performance and predictive power of the MIScnn showed impressive predictions and a very good segmentation performance. We proved that with just a few lines of codes using the MIScnn framework, it was possible to successfully build a powerful pipeline for medical image segmentation. Yet, this performance was achieved using only the standard U-Net model. Fast switching the model to a more precise architecture for high resolution images, like the Dense U-Net model, would probably result into an even better performance [10]. However, this gain would go hand in hand with an increased fitting time and higher GPU memory requirement, which was not possible with our current sharing schedule for GPU hardware. Nevertheless, the possibility of swift switching between models to compare their performance on a data set is a promising step forward in the field of medical image segmentation.

## 4.3 Road Map and Future Direction

The active MIScnn development is currently focused on multiple key features: Adding further data I/O interfaces for the most common medical image formats like DICOM, extend preprocessing and data augmentation methods, implement more efficient patch skipping techniques instead of excluding every blank patch (e.g. denoising patch skipping) and implementation of an open interface for custom preprocessing techniques for specific image types like MRIs. Next to the planned feature implementations, the MIScnn road map includes the model library extension with more state-of-the-art deep learning models for medical image segmentation. Additionally, an objective comparison of the U-Net model version variety is outlined to get more insights on different model performances with the same pipeline. Community contributions in terms of implementations or critique are welcomed and can be included after evaluation. Currently, MIScnn already offers a robust pipeline for medical image segmentation, nonetheless, it will still be regularly updated and extended in the future.

## 4.4 MIScnn Availability

The MIScnn framework can be directly installed as a Python library using pip install miscnn. Additionally, the source code is available in the Git repository: *https://github.com/frankkramer-lab/MIScnn*. MIScnn is licensed under the open-source GNU General Public License Version 3. The code of the cross-validation experiment for the Kidney Tumor Segmentation Challenge is available as a Jupyter Notebook in the official Git repository.

# 5 Conclusion

In this paper, we have introduced the open-source Python library MIScnn: A framework for medical image segmentation with convolutional neural networks and deep learning. The intuitive API allows fast building medical image segmentation pipelines including data I/O, preprocessing, data augmentation, patch-wise analysis, metrics, a library with state-of-the-art deep learning models and model utilization like training, prediction, as well as fully automatic evaluation (e.g. cross-validation). High configurability and multiple open interfaces allow users to fully customize the pipeline. This framework enables researchers to rapidly set up a complete medical image segmentation pipeline by using just a few lines of code. We proved the MIScnn functionality by running an automatic cross-validation on the Kidney Tumor Segmentation Challenge 2019 CT data set resulting into a powerful predictor. We hope that it will help migrating medical image segmentation from the research labs into practical applications.

## Acknowledgments

We want to thank Bernhard Bauer and Fabian Rabe for sharing their GPU hardware (2x Nvidia Quadro P6000) with us which was used for this work. We also want to thank Dennis Klonnek, Florian Auer, Barbara Forastefano, Edmund Müller and Iñaki Soto Rey for their useful comments.

## Conflicts of Interest

The authors declare no conflicts of interest.

## References

[1]     P. Aggarwal, R. Vig, S. Bhadoria, A. C.G.Dethe, Role of Segmentation in Medical Imaging: A Comparative Study, Int. J. Comput. Appl. 29 (2011) 54–61. doi:10.5120/3525-4803.

[2]     Y. Guo, Y. Liu, T. Georgiou, M.S. Lew, A review of semantic segmentation using deep neural networks, Int. J. Multimed. Inf. Retr. 7 (2018) 87–93. doi:10.1007/s13735-017-0141-z.

[3]     S. Muhammad, A. Muhammad, M. Adnan, Q. Muhammad, A. Majdi, M.K. Khan, Medical Image Analysis using Convolutional Neural Networks A Review, J. Med. Syst. (2018) 1–13.

[4]     G. Wang, A perspective on deep imaging, IEEE Access. 4 (2016) 8914–8924. doi:10.1109/ACCESS.2016.2624938.

[5]     G. Litjens, T. Kooi, B.E. Bejnordi, A.A.A. Setio, F. Ciompi, M. Ghafoorian, J.A.W.M. van der Laak, B. van Ginneken, C.I. Sánchez, A survey on deep learning in medical image analysis, Med. Image Anal. 42 (2017) 60–88. doi:10.1016/j.media.2017.07.005.

[6]     D. Shen, G. Wu, H.-I. Suk, Deep Learning in Medical Image Analysis, Annu. Rev. Biomed. Eng. 19 (2017) 221–248. doi:10.1146/annurev-bioeng-071516-044442.

[7]     O. Ronneberger, Philipp Fischer, T. Brox, U-Net: Convolutional Networks for Biomedical Image Segmentation, Lect. Notes Comput. Sci. (Including Subser. Lect. Notes Artif. Intell. Lect. Notes Bioinformatics). 9351 (2015) 234–241. doi:10.1007/978-3-319-24574-4.

[8]     Z. Zhou, M.M.R. Siddiquee, N. Tajbakhsh, J. Liang, UNet++: A Nested U-Net Architecture for Medical Image Segmentation, (2018). http://arxiv.org/abs/1807.10165 (accessed July 19, 2019).

[9]     F. Isensee, J. Petersen, A. Klein, D. Zimmerer, P.F. Jaeger, S. Kohl, J. Wasserthal, G. Koehler, T. Norajitra, S. Wirkert, K.H. Maier-Hein, nnU-Net: Self-adapting Framework for U-Net-Based Medical Image Segmentation, (2018). http://arxiv.org/abs/1809.10486 (accessed July 19, 2019).

[10]    M. Kolařík, R. Burget, V. Uher, K. Říha, M. Dutta, Optimized High Resolution 3D Dense-U-Net Network for Brain and Spine Segmentation, Appl. Sci. 9 (2019) 404. doi:10.3390/app9030404.

[11]    Ö. Çiçek, A. Abdulkadir, S.S. Lienkamp, T. Brox, O. Ronneberger, 3D U-net: Learning dense volumetric segmentation from sparse annotation, Lect. Notes Comput. Sci. (Including Subser. Lect. Notes Artif. Intell. Lect. Notes Bioinformatics). 9901 LNCS (2016) 424–432. doi:10.1007/978-3-319-46723-8_49.

[12]    K. Lee, J. Zung, P. Li, V. Jain, H.S. Seung, Superhuman Accuracy on the SNEMI3D Connectomics Challenge, (2017) 1–11. http://arxiv.org/abs/1706.00120.

[13]    E. Gibson, W. Li, C. Sudre, L. Fidon, D.I. Shakir, G. Wang, Z. Eaton-Rosen, R. Gray, T. Doel, Y. Hu, T. Whyntie, P. Nachev, M. Modat, D.C. Barratt, S. Ourselin, M.J. Cardoso, T. Vercauteren, NiftyNet: a deep-learning platform for medical imaging, Comput. Methods Programs Biomed. 158 (2018) 113–122. doi:10.1016/j.cmpb.2018.01.025.

[14]    Neuroimaging Informatics Technology Initiative, (n.d.). https://nifti.nimh.nih.gov/background (accessed July 19, 2019).

[15]    S. Roy, A. Carass, J.L. Prince, Patch based intensity normalization of brain MR images, in: Proc. - Int. Symp. Biomed. Imaging, 2013. doi:10.1109/ISBI.2013.6556482.

[16]    L.G. Nyú, J.K. Udupa, On standardizing the MR image intensity scale, Magn. Reson. Med. (1999). doi:10.1002/(SICI)1522-2594(199912)42:6<1072::AID-MRM11>3.0.CO;2-M.

[17]    F. Isensee, K.H. Maier-Hein, An attempt at beating the 3D U-Net, (2019) 1–7. http://arxiv.org/abs/1908.02182.

[18]    G. Lin, C. Shen, A. Van Den Hengel, I. Reid, Efficient Piecewise Training of Deep Structured Models for Semantic Segmentation, Proc. IEEE Comput. Soc. Conf. Comput. Vis. Pattern Recognit. 2016-Decem (2016) 3194–3203. doi:10.1109/CVPR.2016.348.

[19]    Z. Hussain, F. Gimenez, D. Yi, D. Rubin, Differential Data Augmentation Techniques for Medical Imaging Classification Tasks., AMIA ... Annu. Symp. Proceedings. AMIA Symp. 2017 (2017) 979–984. https://www.ncbi.nlm.nih.gov/pubmed/29854165.

[20]    Z. Eaton-rosen, F. Bragman, Improving Data Augmentation for Medical Image Segmentation, Midl. (2018) 1–3.

[21]    L. Perez, J. Wang, The Effectiveness of Data Augmentation in Image Classification using Deep Learning, (2017). http://arxiv.org/abs/1712.04621 (accessed July 23, 2019).

[22]    L. Taylor, G. Nitschke, Improving Deep Learning using Generic Data Augmentation, (2017). http://arxiv.org/abs/1708.06020 (accessed July 23, 2019).

[23]    G.C.R.C. (DKFZ) Division of Medical Image Computing, batchgenerators: A framework for data augmentation for 2D and 3D image classification and segmentation, (n.d.). https://github.com/MIC-DKFZ/batchgenerators (accessed October 9, 2019).

[24]    N. Heller, N. Sathianathen, A. Kalapara, E. Walczak, K. Moore, H. Kaluzniak, J. Rosenberg, P. Blake, Z. Rengel, M. Oestreich, J. Dean, M. Tradewell, A. Shah, R. Tejpaul, Z. Edgerton, M. Peterson, S. Raza, S. Regmi, N. Papanikolopoulos, C. Weight, The KiTS19 Challenge Data: 300 Kidney Tumor Cases with Clinical Context, CT Semantic Segmentations, and Surgical Outcomes, (2019). http://arxiv.org/abs/1904.00445 (accessed July 19, 2019).

[25]  P. Coupé, J. V. Manjón, V. Fonov, J. Pruessner, M. Robles, D.L. Collins, Patch-based segmentation using expert priors: Application to hippocampus and ventricle segmentation, Neuroimage. 54 (2011) 940–954. doi:10.1016/j.neuroimage.2010.09.018.

[26]  Z. Zhang, Q. Liu, Y. Wang, Road Extraction by Deep Residual U-Net, IEEE Geosci. Remote Sens. Lett. (2018). doi:10.1109/LGRS.2018.2802944.

[27]  V. Iglovikov, A. Shvets, TernausNet: U-Net with VGG11 Encoder Pre-Trained on ImageNet for Image Segmentation, (2018). http://arxiv.org/abs/1801.05746 (accessed July 19, 2019).

[28]  N. Ibtehaz, M.S. Rahman, MultiResUNet : Rethinking the U-Net Architecture for Multimodal Biomedical Image Segmentation, (2019). http://arxiv.org/abs/1902.04049 (accessed July 19, 2019).

[29]  K. Kamnitsas, W. Bai, E. Ferrante, S. McDonagh, M. Sinclair, N. Pawlowski, M. Rajchl, M. Lee, B. Kainz, D. Rueckert, B. Glocker, Ensembles of multiple models and architectures for robust brain tumour segmentation, Lect. Notes Comput. Sci. (Including Subser. Lect. Notes Artif. Intell. Lect. Notes Bioinformatics). 10670 LNCS (2018) 450–462. doi:10.1007/978-3-319-75238-9_38.

[30]  S. Valverde, M. Salem, M. Cabezas, D. Pareto, J.C. Vilanova, L. Ramió-Torrentà, À. Rovira, J. Salvi, A. Oliver, X. Lladó, One-shot domain adaptation in multiple sclerosis lesion segmentation using convolutional neural networks, NeuroImage Clin. 21 (2019) 101638. doi:10.1016/j.nicl.2018.101638.

[31]  T. Brosch, L.Y.W. Tang, Y. Yoo, D.K.B. Li, A. Traboulsee, R. Tam, Deep 3D Convolutional Encoder Networks With Shortcuts for Multiscale Feature Integration Applied to Multiple Sclerosis Lesion Segmentation, IEEE Trans. Med. Imaging. 35 (2016) 1229–1239. doi:10.1109/TMI.2016.2528821.

[32]  G. Wang, W. Li, S. Ourselin, T. Vercauteren, Automatic brain tumor segmentation using cascaded anisotropic convolutional neural networks, Lect. Notes Comput. Sci. (Including Subser. Lect. Notes Artif. Intell. Lect. Notes Bioinformatics). 10670 LNCS (2018) 178–190. doi:10.1007/978-3-319-75238-9_16.

[33]  Chollet, François, others, Keras, (2015). https://keras.io.

[34]  Martin~Abadi, Ashish~Agarwal, Paul~Barham, Eugene~Brevdo, Zhifeng~Chen, Craig~Citro, Greg~S.~Corrado, Andy~Davis, Jeffrey~Dean, Matthieu~Devin, Sanjay~Ghemawat, Ian~Goodfellow, Andrew~Harp, Geoffrey~Irving, Michael~Isard, Y. Jia, Rafal~Jozefowicz, Lukasz~Kaiser, Manjunath~Kudlur, Josh~Levenberg, Dandelion~Mane, Rajat~Monga, Sherry~Moore, Derek~Murray, Chris~Olah, Mike~Schuster, Jonathon~Shlens, Benoit~Steiner, Ilya~Sutskever, Kunal~Talwar, Paul~Tucker, Vincent~Vanhoucke, Vijay~Vasudevan, Fernanda~Viegas, Oriol~Vinyals, Pete~Warden, Martin~Wattenberg, Martin~Wicke, Yuan~Yu, Xiaoqiang~Zheng, TensorFlow: Large-Scale Machine Learning on Heterogeneous Systems, (2015). https://www.tensorflow.org/.

[35]  S.S.M. Seyed, D. Erdogmus, A. Gholipour, Tversky loss function for image segmentation using 3D fully convolutional deep networks, (2017). https://arxiv.org/abs/1706.05721.

[36]  J. Ferlay, M. Colombet, I. Soerjomataram, C. Mathers, D.M. Parkin, M. Piñeros, A. Znaor, F. Bray, Estimating the global cancer incidence and mortality in 2018: GLOBOCAN sources and methods, Int. J. Cancer. (2018) ijc.31937. doi:10.1002/ijc.31937.